\documentclass[aps,twocolumn,showpacs,nofootinbib]{revtex4}
\usepackage{graphicx}
\usepackage{amssymb}

\usepackage{xcolor}

\begin{document}

\title{Dynamical thermalization in time-dependent Billiards}

\author{Matheus Hansen$^1$, David Ciro$^{2}$, Iber\^e L. Caldas$^{1}$ 
and Edson D.\ Leonel$^{3}$}
\affiliation{$^1$ Instituto de F\'isica - Universidade de S\~ao Paulo, 
S\~ao Paulo - CEP 05508-090  SP, Brazil\\
$^2$ Instituto de Astronomia Geof\'isica e Ci\^encias Atmosf\'ericas - 
Universidade de S\~ao Paulo, S\~ao Paulo - CEP 05508-090 SP, Brazil\\
$^3$ Departamento de F\'{\i}sica - UNESP, Rio Claro - CEP 13506-900 SP, 
Brazil}

\date{\today} \widetext


\begin{abstract}

Numerical experiments of the statistical evolution of an 
ensemble of non-interacting particles in a time-dependent
billiard with inelastic collisions, reveals the existence of three 
statistical regimes for the evolution of the speeds ensemble, namely, diffusion 
plateau, normal growth/exponential decay and stagnation. These regimes 
are linked numerically to the transition from Gauss-like to 
Boltzmann-like speed distributions. Further, the different evolution 
regimes are obtained analytically through velocity-space diffusion 
analysis. From these calculations the asymptotic root mean square of 
speed, initial plateau, and the growth/decay rates for intermediate 
number of collisions are determined in terms of the system parameters. 
The analytical calculations match the numerical experiments 
and point to a dynamical mechanism for \emph{thermalization}, where 
inelastic collisions and a high-dimensional phase space lead to a 
bounded diffusion in the velocity space towards a stationary 
distribution function with a kind of \emph{reservoir temperature} 
determined by the boundary oscillation amplitude and the restitution 
coefficient.
\end{abstract}
\maketitle


\section{Introduction} \label{sec1}

The Loskutov-Ryabov-Akinshin (LRA) conjecture \cite{Ref1} was proposed 
as an attempt to foreseen what would happen to the behavior of the 
average velocity for an ensemble of particles in a time dependent 
billiard \cite{Ref2,Ref3} whenever the shape and characteristics of the 
phase space of the corresponding static version of the billiard was 
known. Chaos in the phase space for the dynamics of a particle in a 
billiard with static boundary was claimed by the conjecture to be a 
sufficient condition to produce unlimited energy growth, also known as 
Fermi acceleration \cite{Ref4}, of the particles when a time 
perturbation to the boundary was introduced. The conjecture was tested 
in a number of billiards being therefore validated \cite{Ref5,Ref6}. A 
counter example of such conjecture was observed in an elliptic billiard 
\cite{Ref7,Ref8}, whose structure is integrable in the static form, but 
that presents an unlimited diffusion of energy when a time-dependence is 
introduced on the billiard boundary.

The physics behind the unlimited energy growth is understood and is 
mainly related to the diffusion of velocities as a function of time 
\cite{Ref9,Ref10,Ref11,Ref12}. Different regimes of growth are related 
to different shapes of the speed distribution function. The counter-intuitive fact
that Hamiltonian dynamics may lead to an unlimited energy growth in chaotic 
billiards comes from the higher dimension of the dynamical system
and such a growth appears to contradict what is expected from 
thermodynamics. However, this only states that there is not a well
defined temperature for the moving boundary, which works here as the energy
reservoir, i.e. the wall is not in a thermodynamic equilibrium.
In a regular situation, a gas of non-interacting particles
with an initial low temperature $T_0$ will increase its energy if 
introduced in a, previously empty, recipient with walls at ambient
temperature $T_a > T_0$. The opposite will happen if the gas is at an
initially larger temperature $T_0>T_a$~\cite{Ref13}. This thermalization
process, in general, manifests as a monotonic change in temperature
as time advances, leading to an asymptotic state of thermal equilibrium.


In contrast, consider a conservative chaotic billiard with
an oscillating boundary, such that unlimited energy growth is observed.
Since the billiard energy is essentially kinetic, the growth of energy leads also 
to the growth of the temperature. The type of interaction of the 
particle with the boundary is the reason of such behavior. Elastic 
collisions preserves both momentum and kinetic energy in the moving 
referential frame of the boundary, which does not imply conservation 
of energy for the inertial frame of the gas center of mass, leading to 
the the unlimited energy growth of the ensemble of particles.

On the other hand, inelastic collisions preserve only momentum, and the 
dissipation introduced produce drastic topological changes in the 
phase space. When inelastic collisions are taken into account, the 
Liouville measure is no longer preserved and attractors can develop
in the phase space \cite{Ref14}. Considering that the 
attractor is located at finite values of the velocity, and its basins 
of attraction contains most of phase space, it is clear that the 
individual trajectories will converge to the attractor, and the average 
speed will saturate leading to a sort of thermodynamical equilibrium 
for the perturbed billiard.

Until now, important results have been obtained in the characterization
of the unbounded energy evolution for particles in chaotic billiards.
Our contribution in this context, is the statistical description of the
evolution to a final equilibrium, and the close connection of this
behavior with thermal equilibrium. To our knowledge this problem has 
not been addressed elsewhere and offers a significant analogy between
dynamical and thermal equilibrium.

In this paper we discuss the dynamics of an ensemble of particles moving 
in an oval billiard with a periodically oscillating boundary. We 
consider inelastic collisions of the particles with the boundary and 
explore the behavior of the root mean square of speed considering the 
shape of the probability distribution function of the speeds. Then we 
show that the presence of dissipation leads the system towards an 
asymptotic stationary state, which, with basis on its statistical 
properties, we argue is a dynamical equivalent of a thermodynamical 
equilibrium. 

The paper is organized as follows. In Sec. \ref{sec2} we 
discuss the equations that compose the billiard model with 
time-dependent boundary, then in Sec. \ref{sec3} we show the statistical 
analysis of the speeds and the diffusion process in the system. In Sec. 
\ref{sec3} we present an analytical derivation of the time evolution of 
the root mean square of the speeds in terms of the control parameters of 
the problem. In Sec. \ref{sec4} we offer a connection between the 
asymptotic dynamics of dissipative time-dependent billiards and the 
concept of thermalization. Finally, in the last section we present our 
conclusions and final remarks.

\section{The time-dependent billiard} 
\label{sec2}

We start considering a time-dependent oval-billiard \cite{Ref15} with 
boundary described in polar form as
\begin{eqnarray}
R_{b}(\theta,\epsilon,t,a,p)&=&1+\epsilon\left[1+a\cos(t)\right]
\cos(p\theta),
\label{eq_02}
\end{eqnarray}
where $R_{b}$ is the boundary radial coordinate, $\theta$ is the polar angle, 
$\epsilon$ measures the oval deformation, $t$ is the time, $a$ is the 
boundary oscillation amplitude and $p$ is a positive 
integer\footnote[1]{Non integer numbers produce open billiard leading 
to escape of particle through hole on the border.}.

\begin{figure}[t]
\centerline{\includegraphics[width=0.78\linewidth]{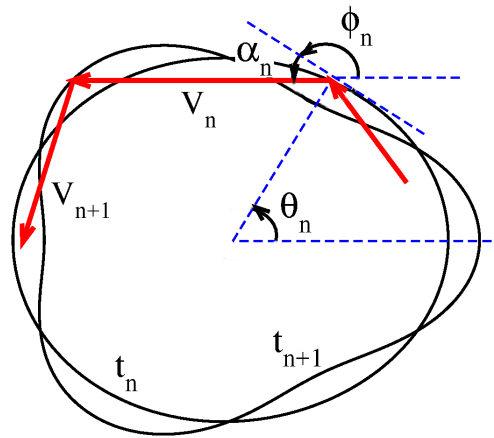}}
\caption{(Color online). Sketch of two consecutive collisions (red 
line) of a particle in a time-dependent oval-billiard with $a=0.9$, 
$\epsilon=0.08$ and $p=3$.}
\label{Fig1}
\end{figure}

The trajectory of a particle inside of the billiard can be described 
using a nonlinear four-dimensional mapping 
$H:\mathbb{R}^4\rightarrow\mathbb{R}^4$, such that 
$(\theta_{n+1},\alpha_{n+1},V_{n+1},t_{n+1})=H(\theta_{n},\alpha_{n},V_{
n},t_{n})$. The angle $\alpha_n$ is measured between the particle 
trajectory and the tangent line to the boundary at $(\theta_n,t_n)$, 
after the $n^{th}$ collision with the wall, and $V_n=|\vec V_n|$ is the velocity 
magnitude. Figure \ref{Fig1} shows in red a sketch of a typical 
trajectory of a particle at different times in the model.

Given that there are no additional potentials inside the billiard, each 
particle moves with constant speed along a straight line between 
collisions. The radial position of the particle is given by 
$R_{p}(t)=\sqrt{X_{p}^{2}(t)+Y_{p}^{2}(t)}$, where $X_p(t)$ and $Y_p(t)$ 
are the rectangular coordinates at time $t$, which are given by
\begin{eqnarray}
X_{p}(t)&=&X(\theta_{n},t_{n})+|\vec{V_{n}}|\cos(\mu)[t-t_{n}], \\
Y_{p}(t)&=&Y(\theta_{n},t_{n})+|\vec{V_{n}}|\sin(\mu)[t-t_{n}], 
\end{eqnarray}
with $\mu=(\alpha_{n}+\phi_{n})$ and 
$\phi=\arctan({Y'(\theta,t)}/{X'(\theta,t)})$, where 
$Y'(\theta,t)=dY/d\theta$ and $X'(\theta,t)=dX/d\theta$. 

The new dynamical variable $\theta$ at collision $n+1$ is obtained 
through the numerical solution of the implicit equation
$R_{b}(\theta_{n+1},t_{n+1})=R_{p}(\theta_{n+1},t_{n+1})$, with the 
time $t_{n+1}$ given by
\begin{eqnarray}
t_{n+1}&=&t_{n}+{{\sqrt{\Delta X_{p}^{2}+\Delta 
Y_{p}^{2}}}\over{|\vec{V_{n}}|}},
\label{eq_03}
\end{eqnarray}
where $\Delta X_{p}=X_{p}(\theta_{n+1},t_{n+1})-X(\theta_{n},t_{n})$ and 
$\Delta Y_{p}=Y_{p}(\theta_{n+1},t_{n+1})-Y(\theta_{n},t_{n})$.

The reflection laws for each collision of the particle with the boundary 
can be obtained by applying conservation of momentum in an instantly 
inertial frame where the contact point of the billiard is at rest. In 
our case, the reflection laws are
\begin{eqnarray}
\vec V_{n+1}'\cdot\vec T_{n+1}&=&\xi\vec V_{n}'\cdot\vec T_{n+1} \nonumber\\
\vec V_{n+1}'\cdot\vec N_{n+1}&=&-\kappa\vec V_{n}'\cdot\vec N_{n+1} \nonumber
\end{eqnarray}
where $\vec T_{n+1}=\cos(\phi_{n+1})\hat{i}+\sin(\phi_{n+1})\hat{j}$ and 
$\vec N_{n+1}=-\sin(\phi_{n+1})\hat{i}+\cos(\phi_{n+1})\hat{j}$ are the 
tangent and normal unit vectors, $\vec V^{'}$ is the particle 
velocity measured in the non-inertial frame, and $\xi,\kappa\in[0,1]$, are 
the tangent and normal restitution coefficients respectively.

After collision $n+1$, the tangent and normal components of the 
velocity are
\begin{eqnarray}
 \vec V_{n+1}\cdot\vec T_{n+1}&=&(1-\xi)\vec V_{b}\cdot\vec T_{n+1}+\xi\vec V_{n}\cdot\vec T_{n+1},\\
 \vec V_{n+1}\cdot\vec N_{n+1}&=&(1+\kappa)\vec V_{b}\cdot\vec N_{n+1}-\kappa\vec V_{n}\cdot\vec N_{n+1},
 \label{eq_04}
\end{eqnarray}
where 
\begin{eqnarray}
\vec V_{b}&=&{dR_{b}(t)\over dt}\Big{|}_{t_{n+1}}[\cos(\theta_{n+1})\hat{i} +\sin(\theta_{n+1})\hat{j}],
\end{eqnarray}
is the boundary velocity at time $t_{n+1}$. The magnitude of the 
particle velocity after collision $n+1$ is 
\begin{eqnarray}
|\vec V_{n+1}|=\sqrt{[\vec V_{n+1}\cdot\vec T_{n+1}]^{2}+[\vec V_{n+1}\cdot\vec N_{n+1}]^{2}},
\label{eq_06}
\end{eqnarray}
and the reflection angle $\alpha_{n+1}$ is 
\begin{eqnarray}
\alpha_{n+1}=\arctan\left[{\vec V_{n+1}\cdot\vec N_{n+1}\over{\vec V_{n+1}\cdot\vec T_{n+1}}}\right].
\label{eq_07}
\end{eqnarray}

\section{System evolution and speed distribution} \label{sec3}

In contrast to the static situation, in a time-dependent billiard, 
particles can gain or lose energy upon interaction with the moving 
boundary. For an ensemble of particles, the individual gains and losses 
do not necessarily compensate and the mean energy can change in time. 
The details of this process can be relevant to understand rates of 
change in the energy but here we start with a simple heuristic analysis
that reveals the broad aspects of the energy evolution.

First of all, consider that the mean quadratic speed changes, by an 
amount $\psi$ after each collision. We consider here a situation in 
which there is an small fractional loss of energy after each collision 
characterized by some restitution coefficient $\gamma<1$, then the mean 
energy after a collision $n$, satisfies 
approximately
\begin{eqnarray}
\overline{V_{n+1}^2}&=&\gamma(\overline{V_n^2}+\psi).
\label{eq_08}
\end{eqnarray}

This dynamical system has a stable equilibrium when the fractional loss 
compensates exactly the energy gain after collision. Regardless of the 
initial configuration, after many collisions, it is expected that the 
quadratic speed $\overline{V_{n+1}^2}$ will approach a stagnation value 
$V_{sta}$ given by
\begin{eqnarray}
V_{sta}&=&\sqrt{{\gamma\psi\over{1-\gamma}}}.
\label{eq_09}
\end{eqnarray}

Notice for elastic collisions $\gamma\rightarrow1$, the stagnation speed 
diverges, which is consistent with the phenomenon of Fermi acceleration, 
where there is an unlimited growth of energy as the time evolves 
\cite{Ref9,Ref12,Ref16}.

To illustrate numerically the stagnation process we consider the root 
mean square of the speed distribution for an ensemble of non-interacting 
particles in the time-dependent oval-billiard described in Section 
\ref{sec1}. To reduce numerical fluctuations, in addition to the 
instantaneous ensemble average, we consider also the time average of the 
quadratic velocity for the ensemble of particles
\begin{eqnarray}
V_{rms}&=&\sqrt{{1\over M}\sum_{i=1}^{M}{1\over{n+1}}\sum_{j=0}^{n}|\vec V|^{2}_{i,j}} \quad,
\label{eq_10}
\end{eqnarray}
where $\vec V_{i,j}$ is the velocity of the $i^{th}$ particle after 
collision $j$. The first summation is made over an ensemble of different 
initial conditions randomly chosen in $t\in[0,2\pi]$, $\alpha\in[0,\pi]$ 
and $\theta\in[0,2\pi]$, where all the particles have the same 
initial speed $V_0$, while the second summation is made over the 
individual orbits. In our simulations we considered an ensemble of 
$M=10^6$ particles colliding $10^7$ times with the boundary.

\begin{figure}[t]
\centerline{\includegraphics[width=\linewidth]{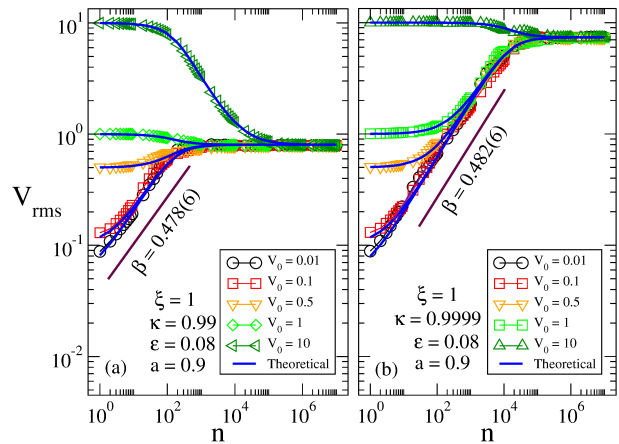}}
\caption{(Color online). (a,b) Plot for $V_{rms}~vs~n$ for different 
initial speeds and parameters.}
\label{Fig2}
\end{figure}

The numerical evolution of $V_{rms}$ is presented Fig. \ref{Fig2}(a,b) 
for two different restitution coefficients $\kappa$, and various initial 
configurations with different $V_0$. These curves in Fig. \ref{Fig2} 
exhibit three different evolution stages for each initial speed. 
Initially, for speeds around of the maximum speed of the boundary 
$V_{max}=a\epsilon$, $V_{rms}$ has a plateau, whose extension 
depends on the initial speed of the particles. After a first crossover, 
the system enters the growth regime following a power law with exponent 
$\beta\sim1/2$ of the number of collisions $n$. Finally, a second 
crossover is observed after which the $V_{rms}$ saturates at $V_{sta}$. 
It can also be observed that when $V_0\gg a\epsilon$, $V_{rms}$ decays 
exponentially to the stagnation regime in agreement with the heuristic 
discussion in Sec. \ref{sec3}.

\begin{figure*}[t]
\centerline{\includegraphics[width=0.7\linewidth]{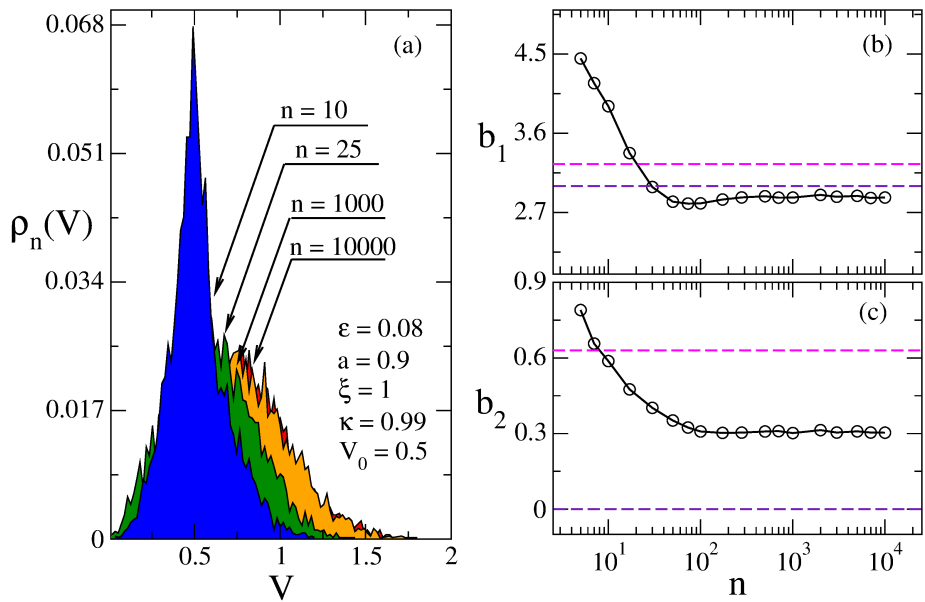}}
\caption{(Color online). (a) Plot of the evolution of the speed 
distribution function $\rho_n(V)$ for an ensemble of $10^6$ particles, 
with initial speed $V_0=0.5$,
after different numbers of collisions $n$ and (b,c) plot of the measurement 
of the kurtosis $b_1$ and skewness $b_2$ (black curve in both figures) respectively, for the 
speed distribution function after different number of collisions. The 
purple and pink dashed lines are the values for the kurtosis and 
skewness for the Normal and two-dimensional Maxwell-Boltzmann 
distributions respectively.}
\label{Fig3}
\end{figure*}

However, to understand in a more detailed fashion the growth rates and 
transition values we need to take into account the diffusion of 
particles in the velocity space $(V_{x},V_{y})$. At the initial stage, 
all particles exist on a circle of radius $V_0$. After a collision, each 
particle jumps by a small amount $(\delta V_x, \delta V_y)$ in some 
direction. Provided there are more available states in the velocity 
space outside the circle than inside, the probability of particles 
moving outside the circle is larger than inwards. This small imbalance 
leads to a growth in the $V_{rms}$ of the ensemble when the initial 
radius $V_0$ is below $V_{sta}$. However, the initial growth rate is 
very small because the initial distribution of particles has to relax 
towards a Gaussian distribution before exhibiting the usual growth rate 
for a random walk $\beta_{RW}=1/2$. Such relaxation process results in 
an initial plateau that is longer for larger initial speeds. For large 
initial velocities $V_0>V_{sta}$, the initial plateau occurs in the same 
fashion, but the probability of moving inwards is larger because with 
each collision the particles must give up an amount of energy 
proportional to $V^2$. Although there are also losses for small 
velocities, the diffusion there dominates because the characteristic value 
of $\delta V^2$ is larger than the energy lost after each collision.

In figure \ref{Fig3}(a), for the same parameters of Fig. \ref{Fig2}(a) and 
$V_0=0.5$, we present the evolution of the speed distribution 
function $\rho_n(V)$ as the number of collisions increases. The 
$10^{th}$ collision corresponds to the plateau region of the Fig. \ref{Fig2}(a) (for $V_0=0.5$) where a spreading 
Gauss-like distribution preserves its mean around the initial speed 
until it reaches $V=0$ at the left side. At the $25^{th}$ collision the 
system is in the growth regime and finally, for the $1000^{th}$ and 
$10000^{th}$ collisions the speed distribution does not change 
appreciably because the system reached its stagnation regime. Thus, comparing 
Fig. \ref{Fig2}(a) and Fig. \ref{Fig3}(a), we can follow, for $V_0=0.5$, the 
$V_{rms}$ and $\rho_n(V)$ evolution as $n$ increases

In order to characterize quantitatively the shape evolution of the speed 
distribution with the number of collisions, we calculated the kurtosis 
$b_1$ and skewness $b_2$ for $\rho_n(V)$ as functions of the number of 
collisions with usual definitions~\cite{Ref17}, given by
\begin{eqnarray}
b_1&=&{1\over M}\sum_{i=1}^M\left[{{V_i-\bar V}\over\sigma}\right]^4,\\
b_2&=&{1\over M}\sum_{i=1}^M\left[{{V_i-\bar V}\over\sigma}\right]^3,
\label{eq_10}
\end{eqnarray}
where $\sigma_{V}=\sqrt{\left<V^2\right>-\left<V\right>^2}$.

Figure \ref{Fig3}(b,c) shows the evolution of $b_1$ and $b_2$ for the 
speed distribution function presented in Fig. \ref{Fig3}(a). This figure 
also shows the values of $b_1$ and $b_2$ for the Normal (purple dashed 
line) and a two-dimensional Maxwell-Boltzmann (pink dashed line) 
distributions, which differ from ours because of the stochastic nature 
of their associated processes. 

Notice that after $\rho_n(V)$ reaches the stagnation regime (about $100$ 
collisions), the kurtosis measurement $(b_1\approx 2.87)$ is close to 
that of a Normal distribution, while the skewness measurement 
$(b_2\approx 0.30)$ is an intermediary value between the Normal and the 
Maxwell-Boltzmann distributions.

\section{Kinetic analysis}\label{sec4}

Consider that the quadratic speed of a single particle, $\tilde V^2$, 
changes by an amount $\psi(\alpha,\theta,t,V)$ after colliding with the 
boundary at position $\theta$ with incidence angle $\alpha$ at time $t$,
\begin{eqnarray}
\tilde V^2(\alpha,\theta,t,V)&=&V^2+\psi(\alpha,\theta,t,V),
\label{eq_11}
\end{eqnarray}
where $V^2$ and $\tilde V^2$ are the quadratic speeds before and after 
collision. In an ensemble of $M$ particles, there are approximately 
$M\mathcal{F}_n(\alpha,\theta,t,V) d\alpha d\theta dt dV$ particles with 
variable $x$ between $x$ and $x+dx$, where $x=\{\alpha,\theta, V, t\}$. 
Then, it is possible to describe the mean quadratic speed after the 
$n^{th}$ collision of an ensemble of particles as
\begin{eqnarray}
\overline{V^{2}_{n+1}}&=&\overline{V^{2}_n}+\delta\overline{V^{2}_{n}},
\label{eq_12}
\end{eqnarray}
where
\begin{figure*}[t]
\centerline{\includegraphics[width=0.9\linewidth]{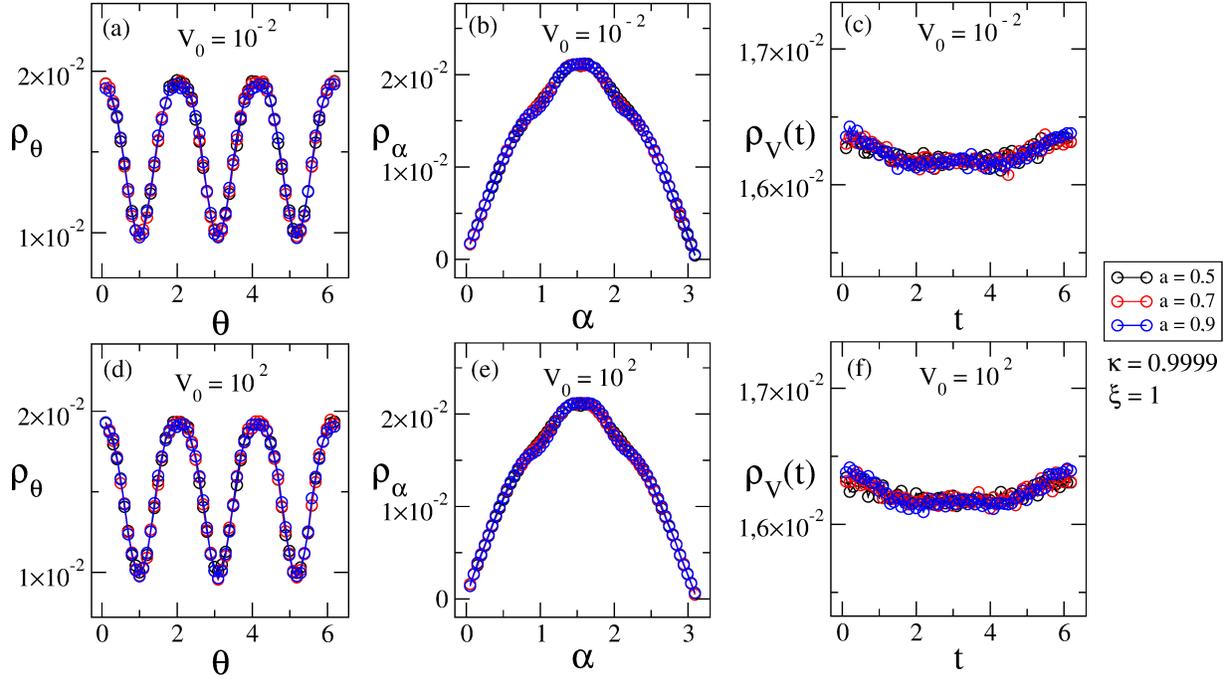}}
\caption{(Color online). (a,b,d,e) Plot of the numerical distribution 
functions $\rho_\theta(\theta)=\int\tilde F(\theta,\alpha)d\alpha$ and 
$\rho_\alpha(\alpha)=\int\tilde F(\theta,\alpha)d\theta$, while (c,d) 
plot of the numerical collision time distribution functions $\rho_V(t)$ 
for the time-dependent oval-billiard at various amplitude of 
oscillations and two initial speeds. The control parameters are 
$\epsilon=0.08$ and $p=3$.}
\label{Fig4}
\end{figure*}
\begin{eqnarray} \label{eq_13}
\overline{V^{2}_{n}} &= \int_0^\infty \!\! \int_0^{2\pi} \!\! \int_0^{2\pi}
\!\! \int_0^\pi& V^2\mathcal{F}_n(\alpha,\theta,t,V) d\alpha d\theta dt dV,\\
\delta\overline{V^{2}_{n}} &= \int_0^\infty \!\! \int_0^{2\pi} \!\! \int_0^{2\pi}
\!\! \int_0^\pi& \psi(\alpha,\theta, t, V)\mathcal{F}_n(\alpha,\theta,t,V)\times \nonumber \\
 && d\alpha d\theta dt dV, 
\end{eqnarray}
and $\mathcal{F}_n(\alpha,\theta,t,V)$ is the full phase space 
distribution function after collision $n$. In our case, we can factor 
this distribution as
\begin{eqnarray}
\mathcal{F}_n(\alpha,\theta,t,V)&=&F(\theta,\alpha)\rho_V(t)\rho_n(V),
\label{eq_14}
\end{eqnarray}
where $F(\alpha,\theta)$ are the angles distribution, $\rho_{V}(t)$ is 
the collision time distribution and $\rho_n(V)$ is the speed 
distribution function.

As observed in Fig. \ref{Fig4}(a-f), the angles $F(\alpha,\theta)$ and 
collision time distributions $\rho_V(t)$, are almost unaffected by the 
amplitude of the boundary oscillations $a$. Consequently they
are almost independent of the index $n$ and on the initial speed values.
This can be understood in terms of the phase-space projection
$\theta-\alpha$, that retains important features of the unperturbed
problem, which is
independent on the velocity of the particles and contains large regular
regions with invariant torii that modulate the size of the chaotic region
as a function of the angles $\theta-\alpha$, in a way consistent with Fig. \ref{Fig4}.

On the other hand, as 
discussed in the previous section, the speed distribution function 
$\rho_n(V)$ (see Fig. \ref{Fig3}(a)) depends on both the speed and the 
index $n$.

Using decomposition (\ref{eq_14}), we want to find an analytical expression that 
describes the behavior of the root mean square speed shown in Fig. 
\ref{Fig2}(a,b). As discussed in \cite{Ref12}, to determine the behavior 
of the mean quadratic speed it is not necessary to describe the evolution 
of the global distribution function, but only to know the evolution of 
its first momenta. Therefore, replacing the Eq. (\ref{eq_14}) in Eq. 
(\ref{eq_13}), and defining the partial mean
\begin{eqnarray}
W(V)\!\! &= \int_0^{2\pi} \!\! \int_0^{2\pi} \!\! \int_0^\pi& \!\! \psi(\alpha,\theta, t, V)
F(\theta,\alpha)\rho_V(t) d\alpha d\theta dt, \qquad
\label{eq_15}
\end{eqnarray}
we can write a more compact expression for the change of the mean 
quadratic speed as
\begin{eqnarray}
\delta\overline{V^{2}_{n}} &= \int_0^\infty& \rho_n(V)W(V)dV.
\label{eq_16}
\end{eqnarray}

As detailed in \cite{Ref12}, if we make a second-order expansion of the 
$W(V)$ around the mean speed $\overline{V_n}$ of the speed distribution 
function $\rho_n(V)$, we obtain 
\begin{equation}
W(V)\approx W(\overline{V_n}) + W'(\overline{V_n})(V-\overline{V_n}) + {1\over{2}} W''(\overline{V_n})(V-\overline{V^{2}_{n}}).
\label{eq_17}
\end{equation}

Inserting the Eq. (\ref{eq_17}) in Eq. (\ref{eq_16}), we obtain an 
approximation for the change of the mean quadratic speed using the 
second-order expansion, as follows
\begin{eqnarray}
\delta\overline{V^{2}_{n}} &=&W(\overline{V_n}) + {1\over{2}} W''(\overline{V_n})(\overline{V}-\overline{V^{2}_{n}}).
\label{eq_18}
\end{eqnarray}

Notice also the Eq. (\ref{eq_17}) is accurate only around the 
distribution mean. As we get far from the mean value the approximation 
becomes poorer. However, $\rho_n(V)$ in the integrand of the Eq. 
(\ref{eq_16}) drops for large and small values of the speed, so the 
integrand is small where the expansion of $W(V)$ is not accurate. The 
interested reader can find more details about these approximations 
in \cite{Ref12}.

Finally, replacing the Eq. (\ref{eq_18}) in Eq. (\ref{eq_12}), we find 
a second-order approximation for the mean quadratic speed after 
collision $n$, as follows
\begin{equation}
\overline{V^{2}_{n+1}} = \overline{V^{2}_{n}}+W(\overline{V_n})+{1\over2}W''(\overline{V_n}(\overline{V^{2}_{n}}-\overline{V^{2}_{n}}).
\label{eq_19}
\end{equation}

In order to determine the partial mean $W(V)$, we first need to find 
$\psi(\alpha,\theta,t)$, which depends on the particular problem. In 
this case, the equations of the time-dependent oval-billiard lead us to
\begin{eqnarray}  \label{eq_20}
 \psi (\alpha,\theta,t) &=& (\kappa^2-1)V^2\sin^2(\alpha)+ \\ \nonumber 
  && +~ (1+\kappa)^2(a\epsilon)^2\cos^2(p\theta)\sin^2(t)+ \\ \nonumber 
  && +~ 2V\kappa a\epsilon(1+\kappa)\sin(\alpha)\cos(p\theta)\sin(t).
\end{eqnarray}

Assuming that the collision time distribution $\rho_V(t)$ is 
approximately uniform, i.e
\begin{eqnarray}
\rho_V(t)&=&{1\over2\pi},
\label{eq_21}
\end{eqnarray}
and replacing the Eq. (\ref{eq_20}) and Eq. (\ref{eq_21}) in Eq. 
(\ref{eq_15}), we obtain
\begin{eqnarray}
 W(V)&=&\eta_1(\kappa^2-1)V^2+{1\over2}(1+\kappa)^2(a\epsilon)^2\eta_2,
 \label{eq_22}
\end{eqnarray}
where
\begin{eqnarray}
\eta_1 &= \int_0^\pi& \sin^2(\alpha) F(\alpha,\theta)d\theta d\alpha,\\
\eta_2 &= \int_0^{2\pi}& \cos^2(p\theta) F(\alpha,\theta)d\theta d\alpha,
\end{eqnarray}
which after inserted in the Eq. (\ref{eq_19}) result in 
\begin{eqnarray}
\overline{V^2_{n+1}}-\overline{V^2_n}=\eta_1(\kappa^2-1)\overline{V^2_n}+{1\over2}(1+\kappa)^2(a\epsilon)^2\eta_2.
 \label{eq_22}
\end{eqnarray}

Considering the approximation of continuous limit 
$\overline{G_{n+1}}-\overline{G_n}\approx dG(n)/dn$, we found a solution 
for the mean quadratic speed
\begin{eqnarray} \label{eq_mean}
 \overline{V^2}&=&\Psi+\left(V_{0}^2-\Psi\right)e^{\eta_1(\kappa^2-1)n},
 \label{quad_ense}
\end{eqnarray}
where
\begin{eqnarray}
 \Psi&=&{(a\epsilon)^2\over2}{\eta_2\over\eta_1}{\left({{1+\kappa}\over{1-\kappa}}\right)}. \nonumber
\end{eqnarray}

To compare with the previous numerical simulations we need to average 
over the ensemble of velocities and the history of velocities of all 
particles. Then we need to average the previous instantaneous mean along 
the history of all quadratic means. i.e 
\begin{eqnarray}
 \overline{V^2}={1\over{n+1}}\sum_{i=0}^{n}\overline{V^2_i}.
 \label{eq_24}
\end{eqnarray}

Provided that the arguments of the exponential are negative, their 
summation converges to
\begin{eqnarray}
 \sum_{i=0}^{n}e^{\eta_1(\kappa^2-1)i}=\left[{1-e^{(n+1)\eta_1(\kappa^2-1)}\over{1-e^{\eta_1(\kappa^2-1)}}}\right].
 \label{eq_25}
\end{eqnarray}

Replacing the Eq. (\ref{eq_25}) in Eq. (\ref{eq_mean}), we obtain 
\begin{eqnarray}
 \overline{V^2}&=&\Psi+\left({{V_0^2-\Psi}\over{n+1}}\right)\left[{1-e^{(n+1)\eta_1(\kappa^2-1)}\over{1-e^{\eta_1(\kappa^2-1)}}}\right].
 \label{eq_26}
\end{eqnarray}

The final expression that describes the root mean square speed 
evolution is 
\begin{eqnarray}
 V_{rms}=\sqrt{\Psi+\left({{V_0^2-\Psi}\over{n+1}}\right)\left[{1-e^{(n+1)\eta_1(\kappa^2-1)}\over{1-e^{\eta_1(\kappa^2-1)}}}\right]},
 \label{eq_28}
\end{eqnarray}
which corresponds to the continuous line (blue) in Fig. 
(\ref{Fig2})(a,b) in excellent agreement with the numerical results for 
the analyzed cases.

To conclude this analytical approach, we consider a few relevant limit 
cases for the Eq. (\ref{eq_28}) that give us relevant insight on the 
overall behavior of the obtained solution. When $n=0$, we have 
\begin{eqnarray}
 V_{rms}&=&V_0,
 \label{eq_29}
\end{eqnarray}
and for $n\rightarrow\infty$ we obtain the finite stagnation value
\begin{eqnarray}
 V_{rms}&=&a\epsilon\sqrt{{1\over2}{\eta_2\over\eta_1}{\left({{1+\kappa}\over{1-\kappa}}\right)}}.
 \label{eq_30}
\end{eqnarray}

\begin{figure*}[t]
\centerline{\includegraphics[width=0.7\linewidth]{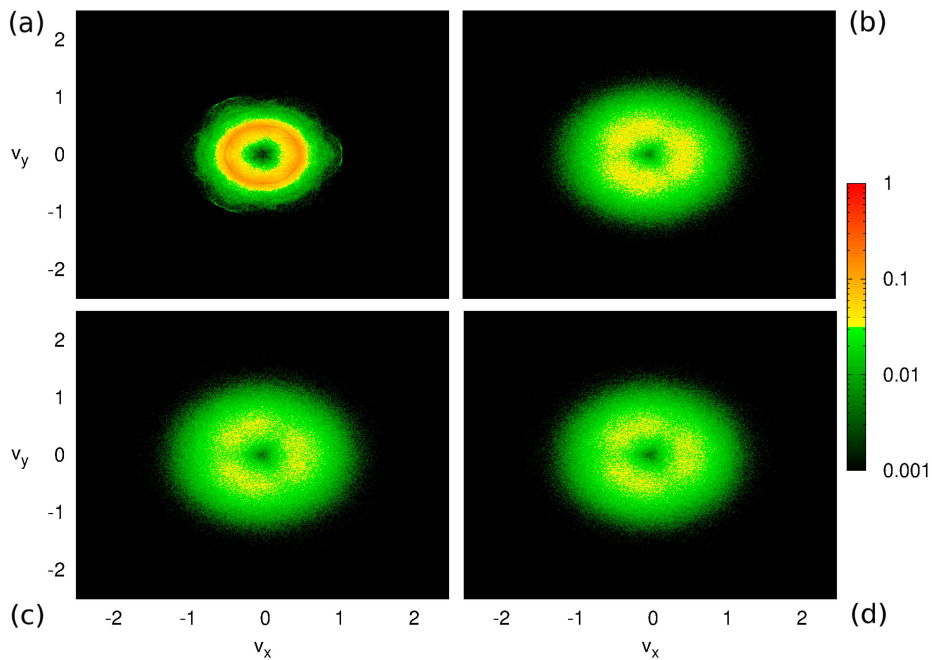}}
\caption{(Color online). Plot of the diffusion speed in the velocities 
space after (a) $10$ collisions; (b) $25$ collisions; 7c) $1000$ 
collisions and (d) $10000$ collisions for an ensemble of $10^6$ 
particles with random positions and locations but with the same initial 
speed $V_0=0.5$. The logarithmic scale of colors represents the density 
of velocity in the velocities space, where the most dense regions are 
shown in red, while the less are in black.}
\label{Fig5}
\end{figure*}

Finally, we consider the intermediate values of $n$ for small initial 
speeds $V_0<<\sqrt{\Psi}$. In the limit of $\kappa\approx1$ we can 
expand to a first order the exponential denominator in the Eq. 
(\ref{eq_28}), while the numerator is taken to the second order due to 
the factor $n+1$ that contributes further to its nonlinearity. After a 
short algebra we obtain 
\begin{eqnarray}
 V_{rms}&\cong&a\epsilon\sqrt{{\eta_2\over2}(1+\kappa)(n+1)},
 \label{eq_31}
\end{eqnarray}
which for $n\gg1$ can be approximated to
\begin{eqnarray}
 V_{rms}&\cong&a\epsilon\sqrt{{\eta_2\over2}(1+\kappa)} n^{1/2}.
 \label{eq_32}
\end{eqnarray}
which leads to the observed growth rate in the numerical treatment of 
the system, and also depends on the boundary oscillation amplitude, the 
angles distribution contained in $\eta_2$, and on the restitution 
coefficient $\kappa$.

\section{Connection with thermodynamics}
\label{sec5}

As a final remark, we present an analogy between dynamical diffusion 
and the stagnation of the mean quadratic speed with the concept of 
thermalization. We start by borrowing the concept of temperature in a 
gas of non-interacting particles inside of a closed region as being 
proportional to the mean of the quadratic peculiar velocities, which, in 
our case is simply the mean of the quadratic velocities because the mean 
velocity of the gas is zero. In this sense, high temperatures are linked 
with high speeds, while the opposite is also true \cite{Ref18}.

We consider the diffusion process in the velocity space 
$(V_{x},V_{y})$ to understand how the ensemble modify with collision. In order to characterize such diffusion, we evolve an 
ensemble of $10^6$ particles in the velocity space, where each one 
started with the same velocity at some point in a circle with radius 
$V_0=0.5$ and random angular position in the billiard.

Figure \ref{Fig5}(a-d) shows the $(V_x,V_y)$ space after $10$, $25$, 
$1000$ and $10000$ collisions, respectively. The color scale measures 
the density of particles. Note that, after $10$ collisions (see Fig. 
\ref{Fig5}(a)) the velocities become distributed around the original 
circle of radius $V_0=0.5$, with concentrated the higher density until 
the particles start populating the velocities near \emph{zero}. This 
behavior is in agreement with the previous discussion of the initial 
plateau discussed in the previous section (see Fig. \ref{Fig2}(a). After $25$ collisions the 
velocities become more spread, and the high concentration circle 
increases its radius, in agreement with the mean speed growth observed 
when the speed distribution becomes asymmetrical. Finally, after $1000$ 
and $10000$ collisions, the distribution does not change much, which, 
expectedly, corresponds to the stagnation state, for which the 
individual velocity fluctuations do not affect the distribution 
function (Fig. \ref{Fig3}(a)).

\begin{figure*}[t]
\centerline{\includegraphics[width=0.6\linewidth]{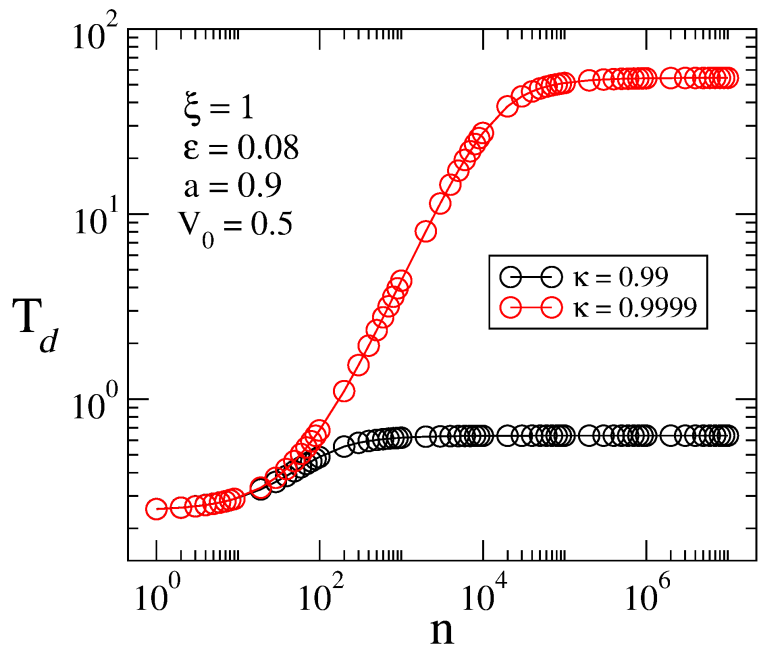}}
\caption{(Color online). Plot of evolution of the {\it dynamical 
temperature} for a gas of non-interacting particles in a time-dependent 
oval billiard in function of the number of collisions.}
\label{Fig6}
\end{figure*}

It is interesting to notice that, the velocity fluctuations are 
responsible to change the mean value of the ensemble of velocities in 
the space $(V_x,V_y)$, where these fluctuations might be estimated as 
the measure of the variance
\begin{eqnarray}
 \sigma_V^2&=&\left<\vec{v}^{~2}\right>-\left<\vec {v}\right>^2,
 \label{eq_33}
\end{eqnarray}
where the mean velocity $\left<\vec{v}\right>$ is \emph{zero} because 
the particles are inside of a non-translating closed billiard, while the 
same does not apply for $\left<\vec{v}^{~2}\right>$, which can be 
identified as the Eq. (\ref{quad_ense}).

Given that we know how is the diffusion process in the system, we can 
define an analogous quantity to the temperature named {\it dynamical 
temperature} $T_d$, that takes into account the characteristics of 
dynamical system studied. This quantity can be written as
\begin{eqnarray}
 T_d&\propto&\overline{V^2}, \\ \nonumber
 \label{eq_34}
\end{eqnarray}
where the equality comes after the introduction of a suitable constant 
$K_d$, which leads us to 
\begin{eqnarray}
 T_d=\frac{m}{2K_d}\left[\Psi+\left(V_{0}^2-\Psi\right)e^{\eta_1(\kappa^2-1)n}\right],
 \label{eq_35}
\end{eqnarray}
with $m$ being the mass of each particle, and $K_d$ plays the role of 
the Boltzmann constant in our dynamical ensemble.

Figure \ref{Fig6} shows the numerical evolution of the {\it dynamical 
temperature} $T_d$ as a function of the number of collisions $n$, for 
two restitution parameters close to \emph{one}. As can be seen, when the 
initial speed $V_0$ is less than $\Psi$, the {\it dynamical temperature} 
of the gas increases with the collisions until it reaches the stagnation 
regime, where it remains for the rest of the simulation. The stagnation 
regime in our context is analogous to a system thermalization with a 
heat reservoir at constant temperature $T_{sta}$ which  emerges from the 
interplay between the oscillating boundary and the inelastic collisions.

Provided there are no additional potentials acting inside of the 
billiard the particles energy is purely kinetic $U_{tot} = E_{k}$, so 
that, we can use the {\it dynamical temperature} to recast the total 
energy in terms of it
\begin{eqnarray}
 U_{tot}&=&E_{k}, \nonumber \\
 U_{tot}&=&{1\over2}N m\overline{V^2}, \nonumber \\
 U_{th}&=&N K_d T_d,
 \label{eq_36}
\end{eqnarray}
where $N$ is the number of particles. Expectedly, due to the definition 
of {\it dynamical temperature}, we recover an energy equation analogous 
to an ideal gas at temperature $T_d$.

\section{Conclusions}

In this work we have described both numerically and analytically the 
development of different evolution regimes for the root mean square of 
speed an ensemble of non-interacting particles in an oscillating 
billiard with collision looses. The analytical treatment was based in 
the diffusion process in the velocity space and resulted in a consistent 
description for all the regimes of the system evolution with 
considerable accuracy. The velocity plateau occurs due to the relaxation 
of the initial configuration, then, for small initial velocities, the 
energy grows with a characteristic exponent close to $1/2$, 
characteristic of normal diffusion otherwise, it will decrease 
exponentially for velocities larger than the saturation speed. Finally, 
the ensemble reaches an stagnation state independent of the initial 
configuration. 

The stagnation regime is analogous to a thermalized 
state, where the distribution function becomes stationary and its 
temperature can be used to characterize the energy reservoir, here 
encompassing the vibrating boundary and the restitution constant. A 
{\it dynamical temperature} was defined to make a connection with 
thermodynamics that resulted in analogous equations for the energy of an 
ideal gas in an actual thermal system.

\section{Acknowledgements}

MH and DC thanks to CAPES for financial support. ILC and EDL thanks to 
CNPq (300632/2010-0 and 303707/2015-1) and S\~ao Paulo Research Foundation 
(FAPESP) (2018/03211-6 and 2017/14414-2) for their financial support.

\end{document}